\documentclass[prl,twocolumn,aps,superscriptaddress,showpacs,floatfix]{revtex4}
\usepackage{graphicx}
\usepackage{dcolumn}
\usepackage{bm}
\usepackage{amsmath}
\usepackage{color}
\begin{document}
\title{Rectification of self-propelled particles by symmetric barriers}
%
\author{A. Pototsky}
\affiliation{Department of Mathematics and Applied Mathematics, University of Cape Town, Cape Town 7701 Rondebosch, South Africa}
\author{A. M. Hahn}
\affiliation{Institut f{\"u}r Theoretische Physik, Technische Universit{\"a}t Berlin, Hardenbergstrasse 36, 10623, Berlin, Germany}
\author{H. Stark}
\affiliation{Institut f{\"u}r Theoretische Physik, Technische Universit{\"a}t Berlin, Hardenbergstrasse 36, 10623, Berlin, Germany}

\begin{abstract}
The motion of self-propelled particles can be rectified by asymmetric or ratchet-like periodic patterns in space.
Here we show that a non-zero average drift can already be induced in a periodic potential with symmetric barriers  when the self-propulsion velocity is also symmetric and periodically modulated but phase-shifted against
the potential. In the adiabatic limit of slow rotational diffusion we determine the mean drift analytically and 
discuss the influence of temperature. In the presence of asymmetric barriers modulating the self-propulsion
can largely enhance the mean drift or even reverse it.

\end{abstract}

\maketitle

Ratchets and rectification of passive Brownian particles or the idea of generating ``noisy transport
far from equilibrium" have been a central topic of statistical physics for the last 20 years with far reaching 
impact on biology and in applications \cite{reimann02,Juelicher97,RMP09}. 
In particular, a detailed analysis under what conditions and symmetries a ratchet allows directed 
transport revealed its basic concepts \cite{reimann02}.

Recent years have seen a tremendously growing interest in understanding self-propulsion in nature 
and of artificial microswimmers \cite{Lauga09,Hong09}.
%
%
Exploring the consequences of self-propulsion currently develops into another paradigm of 
nonequilibrium statistical physics (see, for example, \cite{Toner95,Vicsek95,Narayan07,Tailleur08}).
Also here, rectification of self-propelled particles by static asymmetric barriers 
\cite{austin07,austin08,dileon10,Sokolov10}
has recently attracted much attention due to prospective applications of this easily realizable phenomenon 
in nano- and micro-scale devices. 
Experimental studies with \emph{E. coli} supported by
theoretical work \cite{Tailleur08,wan08,angelani09,dileon11,Kaiser12}
show the pivotal role of self-propulsion for rectifying cell motion in an array of asymmetric funnels 
\cite{austin07,austin08} or for driving a nano-sized ratchet-shaped wheel \cite{dileon10,Sokolov10}. 
Our goal here is to put forward a novel concept of the rectification of active particles, when one is able 
to artificially modulate their self-propulsion velocity. 

Janus particles and their anisotropic surface properties are ideal for generating active motion, for example, 
through self diffusiophoresis in catalytic reaction products \cite{howse07}
%
%
or self thermophoresis under illumination with light \cite{jiang10}. In Refs. \cite{bechinger11,bechinger12} the 
Janus particles are  placed into a critical fluid mixture which allows to fine-tune their self-propulsion 
velocity $v$ by the intensity of the applied light field. In particular, with scanning optical tweezers one 
can then create a periodic modulation of $v$ in space in addition to the presence of a periodically 
corrugated channel or a set of artificially created barriers \cite{bechinger11}.


Motivated by such an experimental system we show in this letter that an arbitray modulation of the self-propulsion
velocity $v(z)$ alone does not generate a mean particle drift. We map the overdamped active Brownian particle moving in a static potential and with a modulated $v(z)$ on a driven pulsating ratchet of a passive particle.
We then demonstrate analytically that in a periodic potential with symmetric barriers active particle motion 
can be rectified with the help of a modulated propulsion velocity. Furthermore, in potentials with asymmetric
barriers, a modulated $v(z)$ can largely enhance the mean particle drift and even reverse it.

We consider 
an active Brownian particle that moves in a spatially patterned environment with a one-dimensional periodic structure characterized by 
a potential $U(z)$
with periodicity $L$. 
In addition, we assume
the propulsion velocity of the particle $v(z)$ 
 to be 
also periodically modulated in space 
along the 
same
direction 
as the periodic pattern. 
More specifically, we set $v(z)=v_0[1 + w(z)]$, 
%
%
where $v_0$ is the 
mean self-propulsion velocity. 
The modulation function $w(z)$ 
has the same periodicity $L$ and satisfies $\int_{z}^{z+L}w(z)\,dz=0$ and
${\rm min}[1 + w(z)]\ge 0$.

 In the overdamped limit, the 
 position
 $\bm{r}=(x,y,z)$ of the 
 active
 particle 
 and its intrinsic direction of self-propulsion given by the unit vector $\bm{p}$
obey the Langevin equations \cite{EncStark11}
\begin{eqnarray}
\label{eq1}
\dot{\bm{r}}=-\mu \partial_z U(z)\bm{e}_z+v(z)\bm{p}+\bm{\xi}(t),\,\,\,\dot{{\bm p}}&=&\bm{\eta}(t)\times \bm{p}.
\end{eqnarray}
%
Here,
$\mu$ 
is the translational mobility and the unit vector $\bm{e}_z$ points 
along the $z$-axis.
$\bm{\xi}(t)$ and $\bm{\eta}(t)$ represent translational and rotational noise, respectively,
with zero mean.
Their variances
are linked to the translational ($\mu k_B T$) and the rotational ($D_r$) diffusivities via the fluctuation-dissipation relations $\langle \bm{\xi}(t) \bm{\xi}(t^\prime) \rangle = 2\mu k_B T \bm{1} \delta(t-t^\prime)$ 
and $\langle \bm{\eta}(t) \bm{\eta}(t^\prime) \rangle = 2D_r \bm{1} \delta(t-t^\prime)$.

According to Eqs.\,(\ref{eq1}), the motion in the $x,y$ plane 
is decoupled from the motion along the $z$-axis. Introducing 
$q = \cos \theta = \bm{p} \cdot \bm{e}_z$,
we can rewrite the equation for the $z$-coordinate of the particle in the form
\begin{eqnarray}
\label{scalar}
\dot{z}&=&v_0q(t)-\frac{\partial U_{\rm eff}(z,q(t))}{\partial z}+\xi(t),
\end{eqnarray}
where we introduced
the $L$-periodic 
and
time-dependent 
effective
potential $U_{\rm eff}(z,q(t))$
\begin{eqnarray}
\label{eff_pot}
U_{\rm eff}(z,q(t))=\mu U(z)+q(t) W(z),
\end{eqnarray}
with $W(z)=-v_0\int w(z)\,dz$.
%
One can show
that the stochastic process $q(t)$ satisfies the Stratonovich equation
\begin{eqnarray}
\label{eff_q}
\dot{q}&=&-D_r q+\sqrt{1-q^2}\eta(t),
\end{eqnarray}
where $\eta(t)$ is a Gaussian white noise with $\langle \eta(t)\eta(t^\prime) \rangle=2D_r\delta(t-t^\prime)$.
The stationary distribution 
$R(q)$,  and the 
time correlation function $\langle q(t)q(t^\prime)\rangle$ are given by
\begin{eqnarray}
\label{stats_3d}
R(q)&=&1/2,\,\,\,\,
\langle q(t)q(t^\prime)\rangle=(1/3)e^{-2D_r\mid t-t^\prime\mid },
\end{eqnarray}
respectively \cite{note}.

 Equation\,(\ref{scalar}) shows that 
 an active Brownian particle 
 moving
in a patterned environment with a spatially modulated propulsion velocity is equivalent to 
a passive Brownian particle 
moving
in a randomly pulsating potential $U_{\rm eff}(z,q(t))$
and
driven by a random force $v_0 q(t)$.
The latter is widely known as a`driven pulsating ratchet'' \cite{RMP09,reimann01}.
Here we have a
special type of 
stochastic pulsation and 
a stochastic driving force $\propto v_0q(t)$. 



Now we investigate under which conditions the active particle system generates a non-zero 
mean drift velocity $\langle \dot{z}\rangle$.
First, using
similar arguments as in Ref.\,\cite{reimann01}, 
we show
that in the absence of the external potential, $U(z)=0$, the average drift of the particle
vanishes exactly for an {\it arbitrary} spatially modulated propulsion velocity $v(z)$ 
meaning $\langle \dot{z}\rangle = 0$. 
We
introduce an auxiliary stochastic process $Z(t)=z(-t)$
with
$\dot{Z}=-dz(-t)/d(-t)$
and, consequently, 
obtain
$\langle \dot{Z}\rangle = -\langle \dot{z}\rangle$. The time evolution equation for $Z(t)$ 
follows
from Eq.\,(\ref{scalar}),
\begin{eqnarray}
\label{scalar_Z}
\dot{Z}&=&-v_0q(-t)-w(Z)q(-t)-\xi(-t).
\end{eqnarray}
We recall
that $-\xi(-t)$ is again a Gaussian white noise.
Also the statistical properties of $-q(-t) = -\bm{e}_z \cdot \bm{p}(-t)$ are identical to
those of $q(t)$ (parity-time invariance \cite{reimann02}) since ${\bm p}(t)$ simply performs a random walk 
on the unit sphere.
Consequently, we can substitute in Eq.\,(\ref{scalar_Z}) $-\xi(-t)$ by $\xi(t)$ and $-q(-t)$ by $q(t)$
and obtain
$\langle \dot{Z}\rangle=\langle\dot{z}\rangle=-\langle\dot{z}\rangle=0$.

Second, in a symmetric potential $U(z)$ the active-particle current is zero for constant self-propulsion velocity.
 However, we now show 
that already
a symmetric velocity modulation $w(z)$ 
can be suffcient to induce a non-zero mean current.
Again, we introduce
an auxiliary variable $Y(t)=-z(t)+z_0$, with some arbitrary 
displacement
$z_0$
and obtain from Eq.\,(\ref{scalar}),
\begin{eqnarray}
\label{scalar_Z2}
\dot{Y}&=&-v_0q(t)-w(-Y(t)+z_0)q(t)\nonumber\\
&+&\mu U^\prime(-Y(t)+z_0)-\xi(t).
\end{eqnarray}
We replace the stochastic processes $-q(t)$ and $-\xi(t)$
in Eq.\,(\ref{scalar_Z2})  by $q(t)$ and $\xi(t)$ using their time-reversal symmetry. We then ensemble 
average $\dot{Y}$ and conclude
that the mean current is always zero, $\langle \dot{Y}\rangle=\langle\dot{z}\rangle=-\langle\dot{z}\rangle=0$,
when a unique displacement $z_0$ exists, for which both potential $U(z)$ and velocity modulation $w(z)$
become symmetric
about $z=0$.
This means $U(-z+z_0) = U(z+z_0)$ and $w(-z+z_0) = w(z+z_0)$.
So, only when the left-right symmetry along the $z$ direction is broken, for example, by a non-zero phase 
shift between $U(z)$ and $w(z)$, can one obtain a non-zero mean current. 
Another possibility would be to introduce a ratchet-like modulation of $w(z)$. We concentrate here
on the first case.

%
%

\emph{Symmetric potential and velocity modulation:}
We now investigate in more detail
the most striking manifestations of the last result
that one can
rectify active swimmers with a symmetrically modulated 
self-propulsion
velocity in a static and symmetric potential.
For the sake of argument, we set $v(z)=v_0[1+a\sin{(kz+\delta)}]$ and $U(z)=U_0\sin{kz}$,
where we introduced the wave number $k=2\pi/L$.
From the above symmetry arguments we expect a non-zero current $\langle \dot{z}\rangle\not=0$ 
for $\delta\not=\pi n,\,n=0,\pm 1,\dots$

In order to unveil the bare physical mechanism of the 
rectification
we will carry out the 
following
analysis 
for
zero temperature, i.e. $\xi(t)=0$ in Eq.\,(\ref{scalar}). 
In addition,
we will assume that the stochastic process $q(t)$ changes adiabatically slow in time. In case of active 
Brownian particles this corresponds to small rotational diffusivities. More precisely, the 
rotational decorrelation time $\propto D_r^{-1}$
should be much larger than the time 
$L/v_0$
needed for the 
active
particle to swim the distance
$L$ of one period of the potential.

\begin{figure}
\centering
\includegraphics[width=0.45\textwidth]{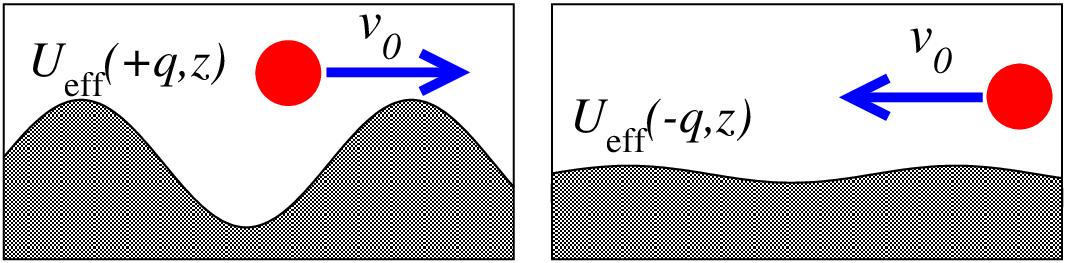}
\caption{(Color online)  Schematic representation of the rectification.
The active particle experiences different effective potentials Eq.\,(\ref{eff_pot}) (shown by shaded area), 
when moving in opposite directions. \label{F0}}
\end{figure}

In Fig.\ref{F0} we illustrate the basic mechanism for the rectification, where we plot the effective potential 
$U_{\mathrm{eff}}$ for particles moving in positive ($+ q > 0$) and negative ($-q$) direction. Clearly,
the particles moving to the right experience steeper potential rises or larger opposing forces, which can even
trap the particle in a mimimum if the force is too large. Whereas in the opposite direction, the potential here
is nearly constant. So, the mean drift velocity  $\langle \dot{z} \rangle$
will be negative. Only for phase differences $\delta = 0, \pi, \ldots$, the shapes of 
$U_{\mathrm{eff}}(+q)$ and $U_{\mathrm{eff}}(-q)$ are identical but phase shifted against each other
and $\langle \dot{z} \rangle = 0$.

With the potential $U(z)$ and the velocity modulation $w(z)$ both sinusoidal, the noiseless 
Eq.\,(\ref{scalar}) can be conveniently written as the well-known Adler equation \cite{adler}
\begin{eqnarray}
\label{adler}
\dot{z}=v_0 q +A(q)\sin{[kz+\phi(q)]},
\end{eqnarray}
with
amplitude
$A(q)$\,$=$\,$\sqrt{(qav_0\cos{\delta})^2+(qav_0\sin{\delta}-k\mu U_0)^2}$ and 
phase shift $\tan{\phi}=(qav_0\sin{\delta}-k\mu U_0)/qav_0\cos{\delta}$
considered as a constant in the adiabatic limit.

The long-time solution of Eq.\,(\ref{adler}) is either a pair of two fixed points in $z\in [0,L]$ for $\vert q\vert v_0 < A(q) $ 
meaning the active particle with $\dot{z}=0$ is trappend in a minimum of $U_{\mathrm{eff}}-v_0qz$.
Or
the time-dependent solution for $\vert q\vert v_0 > A(q)$ 
increases ($q > 0$) or decreases ($q < 0$)
in time.
In fact, the solution for $\vert q \vert v_0> A(q)$ can be decomposed into a periodic component and a dc component representing the average slope $\langle\dot{z}\rangle(q)$ (time averaged $\dot{z}$ at fixed $q$), which is given by
\begin{eqnarray}
\label{dzdt}
\langle \dot{z} \rangle(q)=qv_0\sqrt{1-(A(q)/qv_0)^2}.
\end{eqnarray}

%
%

The average zero temperature (zero-$T$) current $\langle \dot{z}\rangle$ is found by averaging the slope $\langle \dot{z} \rangle(q)$ from Eq.\,(\ref{dzdt}) over all possible $q$ with the stationary distribution function $R(q)=1/2$ from Eq.\,(\ref{stats_3d}). i.e. as $\langle \dot{z}\rangle = \int_{-1}^{+1}\langle \dot{z} \rangle(q)\,R(q)\,dq$.

For arbitrary 
phase shift
$\delta$ and 
amplitude $a$ of the velocity modulation,
$\langle \dot{z}\rangle$
can be computed analytically
but
is given by a rather lengthy algebraic expression. For the sake of clarity, we only 
mention
the zero-$T$ current in two special cases: (i) 
when
$a=1$ and $\delta=\pi/2$ (meaning the self-propulsion is zero at specific points in space), and (ii) 
when $v_0\rightarrow \infty$.
%
%
The corresponding zero-$T$ currents $\langle \dot{z}\rangle^{(i),(ii)}$ are given by
\begin{eqnarray}
\label{current}
\langle \dot{z}\rangle^{(i)} =\frac{\sqrt{k\mu U_0}}{6v_0}(2v_0-k\mu U_0)^{3/2}\,\,\,\,\langle \dot{z}\rangle^{(ii)} =\frac{k\mu a U_0\sin{\delta}}{\sqrt{1-a^2}}.
\end{eqnarray}
From Eqs.\,(\ref{current}) we conclude that the current remains finite in the limit $v_0\rightarrow \infty$ for $\mid a\mid<1$, and that for $a=1$, the current increases unlimited with $v_0$ according to the asymptotic law $\langle \dot{z}\rangle \sim \sqrt{k\mu U_0 v_0}$. 
Furthermore, by
inspecting Eqs.\,(\ref{adler}) and (\ref{dzdt}), we 
find that the optimal rectification regime is reached for $\delta=\pi/2$ and $a=1$.

The system studied so far has striking similarities with the classical {\it Seebeck} ratchet, where a passive
particle moves in a static periodic potential under a periodic temperature profile\ \cite{reimann02}. Typically, 
both fields are also symmetric and have the same periodicity. Only when the spatial symmetry of the
system is broken, \emph{i.e.}, when there is a non-zero phase shift between potential and temperature profile,
does one observe a mean particle velocity exactly as in the case discussed here. 
The profound difference is the term with the self-propulsion velocity $v_0$ in Eq.\ (\ref{scalar}).
Even at zero temperature and in the limit  $D_r \rightarrow 0$, where $q$ only changes adiabatically, 
can one observe a non-zero current of active particles.

For any 
non-zero temperature, the rectification current $\langle \dot{z}\rangle$ of active Brownian particles can be 
found in the adiabatic approximation by using standard results for passive Brownian particles \cite{reimann02}. Thus, for any fixed value of $q=q(t)$, the current $J(q)$ is obtained as 
the
current of a passive particle
that moves in the effective potential Eq.\,(\ref{eff_pot}) 
and is driven by the constant 
effective force $v_0 q$. 
According to \cite{reimann02} we find the stationary conditional density $\rho_s(z|\,q)$ and the current $J(q)$
and give their lengthy formulas explicitely in the supplemental material.
In what follows, we refer to $J(q)$
as the finite-$T$ adiabatic current.
%
%
%
The total rectification current $\langle J \rangle$ 
then becomes $\langle J\rangle = \frac{1}{2} \int_{-1}^{1}J(q)\,dq$.


%
%
%

%
\begin{figure}
\centering
\includegraphics[width=0.49\textwidth]{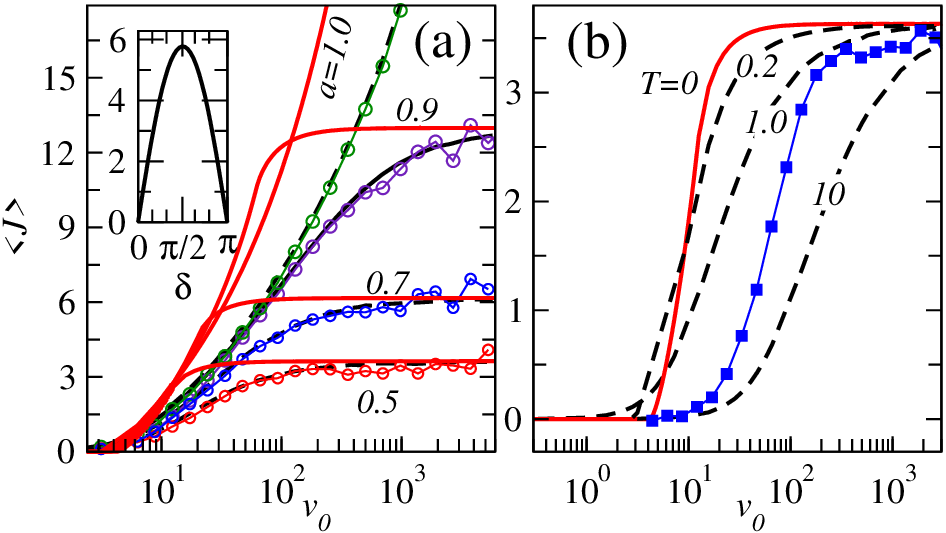}
\caption{(Color online) 
Average current $\langle J \rangle$  in nondimensional units for 
$U(z)=\sin{2\pi z}$ and $w(z)=a\sin{(2\pi z+\delta)}$ with $D_r=20$.
Solid red and dashed lines show the zero-$T$ and finite-$T$ adiabatic limits of $\langle J \rangle$, respectively.
Symbols correspond to numerical solutions of the dimensionless Eqs.\,(\ref{eq1}).
(a) $\langle J \rangle$ {\it vs} $v_0$ for different $a$ and with $\delta=\pi/2$ and $T=1$.
Inset shows the $\delta$ dependence at $a=1$ and $v_0=60$.
(b) $\langle J \rangle$ {\it vs} $v_0$ for different temperatures $T$ at $a=0.5$ and $\delta = \pi/2$. 
Symbols belong to $D_r=400$ and $T=1$.
}
\label{F1}
\end{figure}
%
%
%
%

In order to illustrate our analytical results, we compare the finite-$T$ and the zero-$T$ average current 
$\langle J\rangle$ in the adiabatic limit to numerical results by solving Eqs.\ (\ref{eq1}).
We use reduced units, where we rescale position $z$  by period $L$ and time $t$ by the characteristic 
time $\tau = (\mu U_0/L^2)^{-1}$ which a passive particle needs to travel a distance $L$ when
it is driven by force $U_0/L$. We also refer all energies to $U_0$ so that in particular the
potential amplitude becomes 1. This introduces nondimensional parameters for which we use the
same symbol; namely the reduced mean propulsion velocity $v_0 L/ \mu U_0 \rightarrow v_0$, the 
reduced temperature $k_B T/ U_0 \rightarrow T$, and the rotational diffusivity
$\tau D_r \rightarrow D_r$.

 

In Fig.\ \ref{F1} we show several graphs of the average current $\langle J \rangle$. Solid red and dashed
lines always correspond to the zero-$T$ and finite-$T$ currents in the adiabatic limit. Symbols are obtained 
from direct numerical solutions of the nondimensional Langevin equations\ (\ref{eq1}) with $D_r=20$, implying 
that the adiabatic limit is expected for $v_0 \gtrsim 20$.

%
%
 
In Fig.\,\ref{F1}(a) both
the finite-$T$ and the zero-$T$ currents increase monotonically with $v_0$
and saturate at
the respective limiting value
$\langle \dot{z}\rangle^{(ii)}$ given in 
Eq.\,(\ref{current}) for any $\vert a \vert <1$.
The overall effect of temperature is to reduce the current since it disturbes the directed motion
of the active particles and thereby decreases the difference between particle currents in opposite
directions.
As predicted by $\langle \dot{z}\rangle^{(i)}$ in Eq.\,(\ref{current}),
the zero-$T$ current diverges at $a=1$ as $v_0 \rightarrow \infty$. This implies that the current 
can be made as large as one desires by increasing $v_0$. However, 
the current grows rather weakly since $\langle J \rangle  = \langle \dot{z} \rangle \sim \sqrt{v_0} $


The finite-$T$ current increases monotonically with $a$. Moreover, $\langle J \rangle$ is a $2\pi$-periodic 
function in $\delta$ and symmetric about $\delta = \pi/2$ as the inset of Fig.\,\ref{F1}(a) demonstrates.
So the maximum current occurs at $\delta=\pi/2$.


Figure\,\ref{F1}(b) clearly shows that the zero-$T$ current is zero until $v_0$ reaches a value
of ca.\ $2 \pi$, the maximum force from the static potential $U(z)$.
The active particle needs a non-zero $v_0$ to move against the potential force.
Temperature fluctuations help the particle to cross potential barriers and thereby activate a non-zero current.
As already mentioned, for $v_0 > 2\pi$ temperature fluctuations smear out the difference between the 
active motion in opposite directions, and, as a consequence, the zero-$T$ current is, generally, larger than 
the finite-$T$ current.

Interestingly, large temperatures such as $T=10$ in Fig.\,\ref{F1}(b) no longer initiate a non-zero current 
at small $v_0$ since they dominate the active-particle motion completely.
Furthermore, the largest rectification current is generally achieved in the adiabatic limit since
rotational diffusion merely diminishes the current as demonstrated in Fig.\ref{F1}(b) with numerical
results at $D_r=400$.

%
%

\begin{figure}[ht]
\centering
\includegraphics[width=0.49\textwidth]{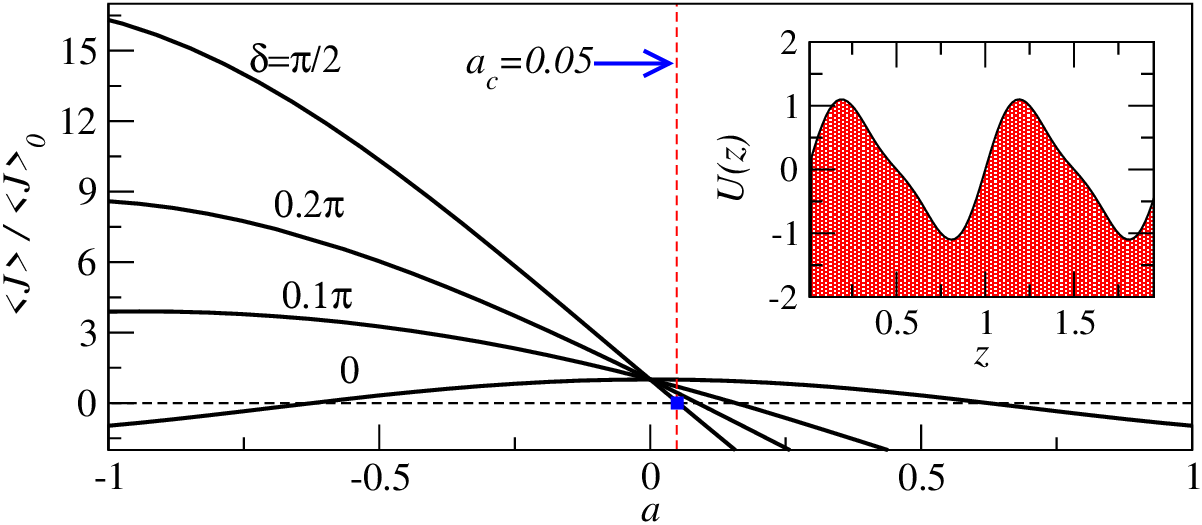}
\caption{(Color online) Finite-$T$ average current $\langle J \rangle$ in units of 
$\langle J \rangle_0=\langle J \rangle(a=0)$ 
as a function of $a$ in case of 
$U=\sin(2\pi z)+(1/4)\sin{4\pi z}$
and symmetric velocity modulation $w(z)$ as in Fig.\,\ref{F1}. 
Other parameters are $v_0=10$, $T=1$, 
and different $\delta$ as indicated. Current inversion at $\delta=\pi/2$ occurs at $a=a_c=0.05$, as indicated by the vertical dashed line and a filled square. Inset shows $U(z)$.
\label{F2}
}
\end{figure}

 \emph{Strong current enhancement and reversal:} 
Now, we demonstrate that a weak modulation of the 
self-propulsion velocity can have a pronounced effect on the average active current, which is induced by 
a static asymmetric or ratchet-like potential. The modulation either very strongly enhances the 
current or even reverses it.
To this end we set 
$U(z)=\sin{2\pi z}+(1/4) \sin{4\pi z}$
(see inset of Fig.\,\ref{F2})
and use the same sinusoidal velocity modulation function $w(z)$ as before.

In Fig.\,\ref{F2} 
we plot $\langle J \rangle$ versus the
amplitude of the velocity modulation $a$, which we obtained
in the finite-$T$ adiabatic limit for $v_0=10$
which is just above the maximum slope of the static potential, 
$\mathrm{max}(dU(z)/dz)= 9.3$.
%
%
We scale the current
in units of the reference current 
$\langle J\rangle(a=0)$ when the self-propulsion velocity is not modulated.
Different values of the 
phase shift parameter $\delta$ of the 
velocity modulation function
$w(z)$ are used.

%
%
%

%

The strongest effect of the velocity modulation $w(z)$ on the average current is found for $\delta=\pi/2$
and when $a=-1$ meaning that the self-propulsion vanishes at specific points in the potential and
doubles in between. In this case, the current is enhanced by more than an order of magnitude.
Much more sensitive with respect to the modulation amplitude $a$ is the sign of the average current. We find current reversal at any value of the shift parameter $\delta$. Remarkably, for $\delta=\pi/2$, the current is reversed by varying the propulsion velocity by as much as $5\%$, i.e. $a=0.05$, as shown by the red line in Fig.\,\ref{F2}.
%
%
%
\\

Controlling the motion of active or self-propelled Brownian particles is of immense interest both 
for technologial applications but also for a fundamental understanding. Here we 
explored the possibility of rectifying their mean current with the help of a spatially modulated propulsion
speed as realizable, for example, in the experiments of Refs.\ \cite{bechinger11,bechinger12}.
Whereas in a static potential with symmetric barriers the mean current is zero, it becomes non-zero and
can efficiently tuned with a spatially modulated self-propulsion and a phase shift against the
potential. This is reminiscent to the well-known Seebeck ratchets \cite{butt87} of passive systems.
We were able to analytically determine the mean current in the adiabatic limit of slow rotational diffusion for 
any temperature and compared it to numerical results. Finally we demonstrated that the modulated self propulsion
can have a profound effect on the mean current in a ratchet potential. It largely enhances the current or
can even reverse it with only a weak modulation.

The rectification discussed here is universal and does not crucially depend on temperature. 
Therefore, it should also be 
found in non-Brownian active agents, such as the run-and-tumble particles of Ref.\ \cite{tailleur09}.
Our results should be testable with the active Janus particles of Refs.\ \cite{bechinger11,bechinger12},
where the self-propulsion speed can be tuned by light intensity. In particular, the adiabatic limit of slow
rotation mainly used in this article is accessible in the experiments. The rotational decorrelation
time $D_r^{-1}$ due to thermal diffusion depends on the third power of the particle radius and
can be varied in the range $D_r^{-1} = 1 \dots 100 \mathrm{s}$. Together with achievable velocities of 
$v_0 =1 \mathrm{\mu m/s}$ and typical spatial periods of $L\approx 30 \mathrm{\mu m}$
\cite{bechinger11}, the adiabatic limit condition $D_r^{-1}=100 \mathrm{s} \gg L / v_0 = 30 \mathrm{s}$ is 
easily met. This makes the system ideal for exploring the control and rectification of active particles
with modulated self-propulsion speed.

 We thank C. Bechinger and V. Volpe for helpful discussions. This work was partly supported by the research training group GRK 1558, funded by
Deutsche Forschungsgemeinschaft.

\end{document}


\title{Supplementary material}
%
\author{A. Pototsky}
\affiliation{Department of Mathematics and Applied Mathematics, University of Cape Town, Cape Town 7701 Rondebosch, South Africa}
\author{A. Hahn}
\affiliation{Institut f{\"u}r Theoretische Physik, Technische Universit{\"a}t Berlin, Hardenbergstrasse 36, 10623, Berlin, Germany}
\author{H. Stark}
\affiliation{Institut f{\"u}r Theoretische Physik, Technische Universit{\"a}t Berlin, Hardenbergstrasse 36, 10623, Berlin, Germany}


\maketitle
\section{Effective Langevin dynamics in the 3D case}
\label{appendix}
 In the overdamped limit, the coordinate $\bm{r}=(x,y,z)$ of the particle obeys the Langevin equation \cite{EncStark11}
\begin{eqnarray}
\label{eq1}
\dot{\bm{r}}=-\mu \partial_z U(z)\bm{e}_z+v_0(z)\bm{p}+\bm{\xi}(t),\,\,\,\dot{{\bm p}}&=&\bm{\eta}(t)\times \bm{p},
\end{eqnarray}
where $\mu$ stands for the translational mobility, $\bm{e}_z$ is the unit vector of the $z$-axis and the unit vector $\bm{p}$ points along the instantaneous direction of the ballistic motion. $\bm{\xi}(t)$ and $\bm{\eta}(t)$ represent translational and rotational noise, respectively. The later are linked to the absolute temperature $T$ and the rotational diffusivity $D_r$ via the fluctuation-dissipation relations $\langle \bm{\xi}(t) \bm{\xi}(t^\prime) \rangle = 2\mu k_B T\delta(t-t^\prime)$ and $\langle \bm{\eta}(t) \bm{\eta}(t^\prime) \rangle = 2D_r\delta(t-t^\prime)$.

If the multiplicative noise in the second equation in Eqs.\,(\ref{eq1}) is treated in the Stratonovich interpretation, the corresponding Smoluchowski equation for the probability density $\rho(\bm{r},\bm{p},t)$ becomes
\begin{eqnarray}
\label{eq2}
\frac{\partial \rho}{\partial t}+{\rm div} \bm{J}_t = D_r \bm{R}^2 \rho, 
\end{eqnarray}
with the translational current $\bm{J}_t=-\mu\partial_z U(z)\rho\bm{e}_z +v_0(z)\rho\bm{p} - \mu k_B T \bm{\nabla} \rho$ and the rotational operator $\bm{R}=\bm{p}\times \bm{\nabla_p}$ \cite{EncStark11}.

It can be readily shown that 
\begin{eqnarray}
\label{R2}
\bm{R}^2=\bm{\nabla}_p \left( \bm{p}^2 \bm{\nabla}_p -\bm{p} (\bm{p}\cdot \bm{\nabla}_p)\right).
\end{eqnarray}
Taking into account the one-dimensional nature of the modulation function $v_0(z)$ and of the potential $U(z)$, and by using the spherical coordinates for the direction vector $\bm{p}=(\sin{\theta}\cos{\alpha},\sin{\theta}\sin{\alpha},\cos{\theta})$ with $\theta$ representing the angle between the $z$-axis and the unit vector $\bm{p}$, the Smoluchowsky equation Eq.\,(\ref{eq2}) can be written in terms of the distribution density that only depends on one spatial coordinate $\rho(z,\theta,\alpha,t)$
\begin{eqnarray}
\label{eq2_a}
\frac{\partial \rho}{\partial t}+\frac{\partial J_z}{\partial z} = \frac{D_r}{\sin{\theta}}\frac{\partial}{\partial \theta}\left(\sin{\theta}\frac{\partial \rho}{\partial \theta}\right)+\frac{D_r}{\sin^2{\theta}}\frac{\partial^2 \rho}{\partial \alpha^2},
\end{eqnarray}
with the $z$-component of the translational current $J_z=[-\mu \partial_z U(z)+v_0(z)\cos{\theta} - \mu k_B T \partial_z] \rho(z,\theta,\alpha,t)$.

The Smoluchowski Eq.\,(\ref{eq2_a}) can be integrated over the polar angle $\alpha$ to yield
\begin{eqnarray}
\label{eq2_3d}
\frac{\partial \rho}{\partial t}+\frac{\partial J_z}{\partial z} = \frac{D_r}{\sin{\theta}}\frac{\partial}{\partial \theta}\left(\sin{\theta}\frac{\partial \rho}{\partial \theta}\right).
\end{eqnarray}

With the change of variables $q=\cos{\theta}$, the Smoluchowski Eq.\,(\ref{eq2_3d}) is transformed to
\begin{eqnarray}
\label{app_1}
&&\frac{\partial \rho}{\partial t}+\frac{\partial J_z}{\partial z} = D_r\frac{\partial}{\partial q}\left((1-q^2)\frac{\partial \rho}{\partial q}\right)\nonumber\\
&=&\frac{\partial}{\partial q}\left(D_r q\rho +D_r\sqrt{1-q^2}\frac{\partial}{\partial q}\left[\sqrt{1-q^2}\rho\right]\right),
\end{eqnarray}
with $\rho=\rho(z,q,t)$.

The expression in the round brackets in the r.h.s. of Eq.\,(\ref{app_1}) represents the (negative) probability current $j_q$ of the stochastic process $q(t)$ 
\begin{eqnarray}
\label{app_2}
-j_q=D_r q\rho +D_r\sqrt{1-q^2}\frac{\partial}{\partial q}\left[\sqrt{1-q^2}\rho\right],
\end{eqnarray}
which corresponds to a Langevin equation in the Stratonovich interpretation
\begin{eqnarray}
\label{app_3}
\dot{q}=-D_r q +\sqrt{1-q^2}\eta(t),
\end{eqnarray}
where $\eta(t)$ is the white Gaussian noise with the correlation function $\langle \eta(t)\eta(t^\prime)\rangle = 2D_r \delta(t-t^\prime)$.

The stationary distribution $R(q)$ is found by setting the probability current to zero, i.e. $j_q=0$. This yields
\begin{eqnarray}
\label{app_4}
R(q)=const=\frac{1}{2},\,\,\,\,q\in[-1,1].
\end{eqnarray}
In order to compute the conditional average $\langle q(t)\mid q_0,t_0\rangle$, we use the time-dependent Fokker-Planck equation for the process $q(t)$, which is obtained from the Smoluchowski equation Eq.\,(\ref{app_1}) by integrating the later over the coordinate $z$ in the limits $z\in(-\infty,\infty)$
\begin{eqnarray}
\label{app_5}
\frac{\partial \rho}{\partial t}+\frac{\partial j_q}{\partial q}=0.
\end{eqnarray}
Thus, multiplying Eq.\,(\ref{app_5}) with $u$ and integrating by parts, we have
\begin{eqnarray}
\label{app_6}
\frac{\partial}{\partial t}\langle q(t)\mid q_0,t_0\rangle=\int_{-1}^{1}q\frac{\partial \rho(q,t)}{\partial t}\,dq=\int_{-1}^{1}j_q\,dq,
\end{eqnarray}
where we have used the fact that the time-dependent probability current $j_q$ vanishes at $q=\pm 1$.
Using Eq.\,(\ref{app_2}) we obtain
\begin{eqnarray}
\label{app_7}
\frac{\partial}{\partial t}\langle q(t)\mid q_0,t_0\rangle&=&-D_r \langle q(t)\mid q_0,t_0\rangle \nonumber\\
&-& D_r \int_{-1}^{1}\sqrt{1-q^2}\frac{\partial}{\partial q}\sqrt{1-q^2}\rho\,dq\nonumber\\
&=&-D_r \int_{-1}^{1}(1-q^2)\frac{\partial \rho}{\partial q}\,dq\nonumber\\
&=&-2D_r \langle q(t)\mid q_0,t_0\rangle.
\end{eqnarray}
The general solution of the Eq.\,(\ref{app_7}) is given by
\begin{eqnarray}
\label{app_8}
\langle q(t)\mid q_0,t_0\rangle=q_0e^{-2D_r(t-t_0)}.
\end{eqnarray}
This results can now be used to compute the stationary auto-correlation function (ACF) $\langle q(t)q(t^\prime)\rangle$. It is well known that for a stationary Markov process, the ACF is given by \cite{Gard}
\begin{eqnarray}
\label{app_9}
\langle q(t)q(t^\prime)\rangle &=& \int_{-1}^{1}\langle q(t)\mid q_0,t^\prime\rangle q_0 R(q_0)\,dq_0\nonumber\\
&=&\frac{1}{3}e^{-2D_r(t-t^\prime)},
\end{eqnarray}
where the factor $1/3$ before the exponent represents the stationary variance $\langle q_0^2\rangle_s=(1/2)\int_{-1}^{1}q_0^2\,dq_0=1/3$.
\section{Finite-$T$ adiabatic current}
In the limit of slow rotation, i.e. when $q(t)$ changes slowly with time, we set $q(t)=q=const$ in the equation
\begin{eqnarray}
\label{scalar}
\dot{z}&=&v_0q(t)-\frac{\partial U_{\rm eff}(z,q(t))}{\partial z}+\xi(t),
\end{eqnarray}
and find according to \cite{reimann02} the stationary conditional density $\rho_s(z|\,q)$ and the corresponding current $J(q)$
\begin{eqnarray}
\label{stats}
\rho_s(z|\,q)&=&\frac{J(q)}{\mu k_B T}E^{-1}(z)\left(\frac{I^{+}(L)}{1-e^{-v_0Lq/\mu k_B T}}-I^{+}(z,q)\right),\nonumber\\
J(q)&=&\mu k_B T\left[\frac{I^{+}(L)I^{-}(L)}{1-e^{-v_0Lq/\mu k_B T}}-I(q)\right]^{-1},
\end{eqnarray}
with $I^{\pm}(z)=\int_{0}^{z}E^{\pm 1}(y)\,dy$ and $I(q)=\int_{0}^{L}E^{-1}(z)I^{+}(z)\,dz$ and $E(z)=\exp{\left(\frac{U_{\rm eff}(z,q)-v_0zq}{\mu k_B T}\right)}$.
%